\documentclass[nofootinbib, amsmath,amssymb, aps, prd]{revtex4-2}

\usepackage{graphicx}
\usepackage{dcolumn}
\usepackage{bm}
\usepackage{amsthm}
\newtheorem{prop}{Proposition}
\newtheorem{corr}{Corollary}
\usepackage{braket}

\begin{document}

\preprint{APS/123-QED}


\title{Correlation measures and distillable entanglement in AdS/CFT}


\author{Joshua Levin}
\email{joshua.t.levin@colorado.edu}
\author{Oliver DeWolfe}
\email{oliver.dewolfe@colorado.edu}
\author{Graeme Smith}
\email{graeme.smith@colorado.edu}
\affiliation{Department of Physics and Center for Theory of Quantum Matter, University of Colorado, Boulder CO 80309, USA}

\date{\today}

\begin{abstract}
  Recent developments have exposed close connections between quantum information and holography.  In this paper, we explore the geometrical interpretations of the recently introduced $Q$-correlation and $R$-correlation, $E_Q$ and $E_R$.  We find that $E_Q$ admits a natural geometric interpretation via the surface-state correspondence: it is a minimal mutual information between a surface region $A$ and a cross-section of $A$'s entanglement wedge with $B$.  We note a strict trade-off between this minimal mutual information and the symmetric side-channel assisted distillable entanglement from the environment $E$ to $A$, $I^{ss}(E\rangle A)$.  We also show that the $R$-correlation, $E_R$, coincides holographically with the entanglement wedge cross-section. This further elucidates the intricate relationship between entanglement, correlations, and geometry in holographic field theories.
\end{abstract}

\maketitle


\section{Introduction}

Information measures quantify the various types of information and disorder contained in the state of a quantum system. Familiar information measures such as the mutual information are built as linear combinations of entropies; others, such as the squashed entanglement \cite{CW04}, involve the optimization over additional auxiliary degrees of freedom. A particularly useful class of information measures, which we refer to as {\em correlation measures}, are those that are monotonically decreasing under local quantum operations. Such a monotonic measure captures properties of the state that individual parties cannot create on their own, and thus quantifies correlations between parties. 

The entanglement of purification is an especially interesting information measure \cite{EP02}. It is  defined as a minimization over purifications of the state $\rho_{AB}$ to $\ket{\psi}_{AaBb}$ of the entropy $S(Aa)$,
\begin{align}\label{Eq:EPDef}
    E_P(A:B) = \inf_{\ket{\psi}_{AaBb}}S(Aa),
\end{align}
and can be understood operationally as the entanglement cost of creating the state $\rho_{AB}$ with asymptotically vanishing communication between parties.  The entanglement of purification is monotonic under quantum operations on either party \cite{EP02,Levin19}, and is thus a correlation measure. In \cite{Levin19}, two similar two-party correlation measures were identified: the $Q$-correlation $E_Q(A:B)$ and the $R$-correlation $E_R(A:B)$, defined below. In addition to monotonicity all three of these correlation measures satisfy the inequalities
\begin{align}
\label{Inequalities}
    {1 \over 2} I(A:B) \leq E_\alpha(A:B) \leq {\rm min} (S(A), S(B))\,, \quad \quad E_\alpha (A:BC) \geq {1 \over 2} (I(A:B) + I(A:C))\,,
\end{align}
for $\alpha = P, Q, R$, where $I(A:B) = S(A)+S(B)-S(AB)$ is the mutual information. It is of considerable interest to better characterize the new correlation measures, their operational significance, and their relationship to the entanglement of purification.

In recent years, powerful new tools for exploring quantum information have arisen using holography. The AdS/CFT correspondence relates $d$-dimensional conformal field theories (CFTs) to theories of gravity living in asymptotically anti-de Sitter (AdS) spacetimes in $d+1$ dimensions, where the CFT is thought of as living at the boundary of AdS. In AdS/CFT, the entropy of a spatial region $A$ in the CFT can be calculated in terms of the gravity dual using the Ryu-Takayanagi (RT) formula \cite{RT}, which states that the entropy is proportional to the area of the minimal bulk surface $\Gamma_A$ living in a Cauchy slice $\Sigma$ of AdS homologous to $A$, whose boundary is $\partial A$:\footnote{Here we restrict to time-independent states where this definition is sufficient.}
\begin{align}
S(A) = \frac{{\rm Area}(\Gamma_A)}{4G_N}\,,
\end{align}
where $G_N$ is the gravitational constant.
Inequalities satisfied by information measures such as weak and strong subadditivity can be shown to follow geometrically from the RT formula.
The region in $\Sigma$ bounded by $A$ and $\Gamma_A$ is called the {\em entanglement wedge} of $A$, $EW(A)$.

\begin{figure}
    \centering
    \includegraphics[scale = 0.4, trim = 0 165 0 160, clip]{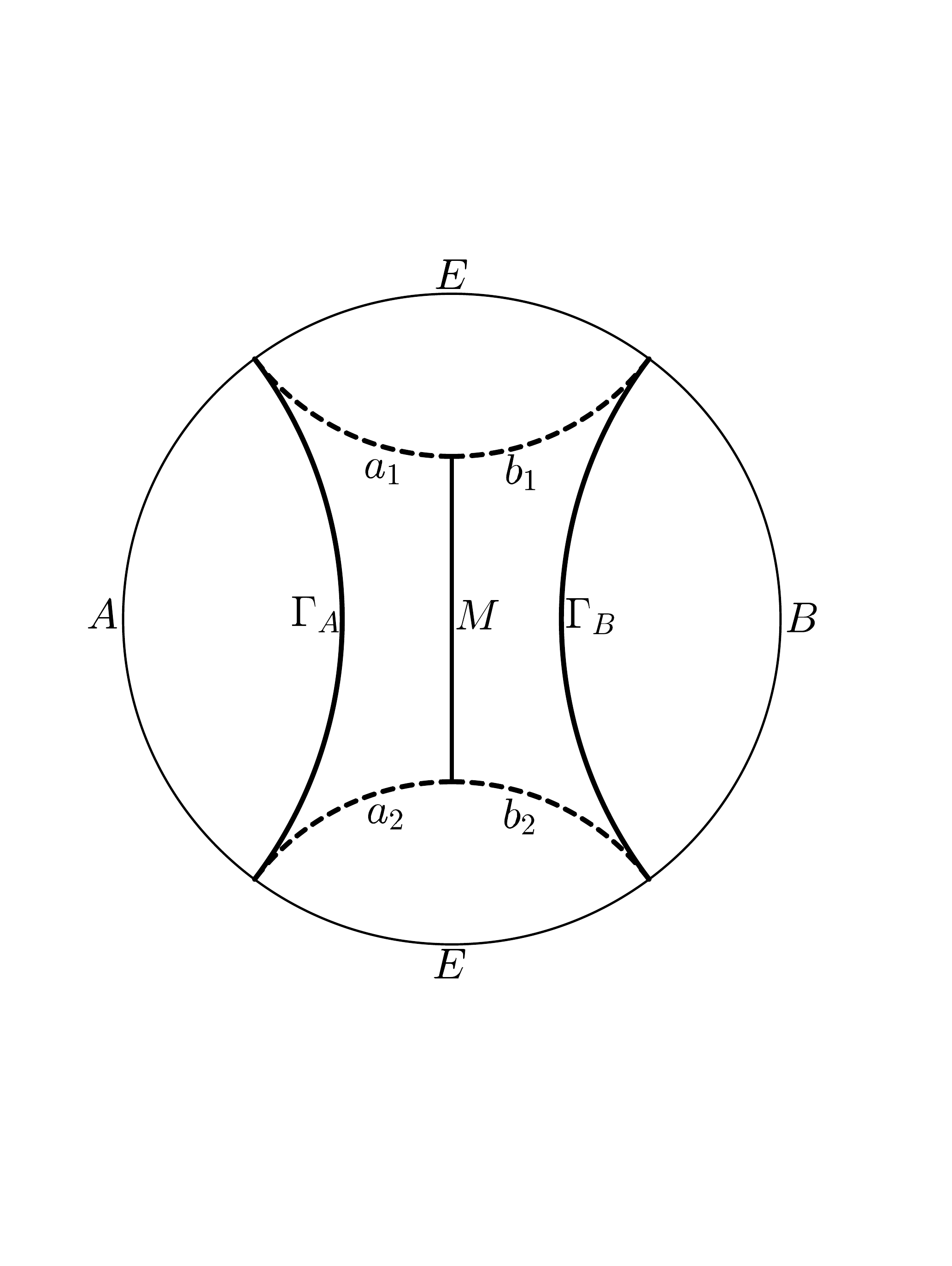}
    \caption{The geometry of the entanglement wedge of $\rho_{AB}$.  $\Gamma_A$ and $\Gamma_B$ are the RT-surfaces of $A$ and $B$, respectively, $M$ is the entanglement wedge cross-section, and $a_1\cup a_2\cup b_1\cup b_2$ is the RT-surface of $AB$.}
    \label{Tennis_ball}
\end{figure}

It is natural to ask whether correlation measures involving the optimization over ancillary degrees of freedom such as $E_P$, $E_Q$ and $E_R$ have a holographic presentation. A proposal for the entanglement of purification was made in \cite{TU18,Swing18}, as follows. Consider two disjoint, connected spatial CFT regions $A$ and $B$. For the sake of concreteness we will focus on a spatial slice in an asymptotically AdS$_3$ geometry dual to a state in a two-dimensional CFT as depicted in Fig.~\ref{Tennis_ball}, where ``surfaces" and ``curves" are interchangeable, though this can be suitably generalized to higher dimensions. For sufficiently small $A$ and $B$ one has that $S(AB) = S(A) + S(B)$ (with RT surfaces depicted by $\Gamma_A$ and $\Gamma_B$ in  Fig.~\ref{Tennis_ball}), and hence the entanglement wedge $EW(AB)$ is the sum of the disconnected components $EW(A)$ and $EW(B)$. On the other hand, for sufficiently large $A$ and $B$ the minimal surface giving $S(AB)$ becomes the dashed curves in  Fig.~\ref{Tennis_ball}, leading to a single, connected entanglement wedge spanning the space from $A$ to $B$.

If $EW(AB)$ is connected then we can define a quantity called the {\em entanglement wedge cross-section}, denoted $E_W$, as the minimum of the area of a surface partitioning $EW(AB)$ into two regions, one containing $A$ in its boundary and the other containing $B$. Denote this minimal surface by $EWCS(AB)$. The surfaces bounding the entanglement wedge are necessarily geodesics, and it is a fact of 2D hyperbolic geometry that any two complete parallel geodesics (which do not meet at infinity) have a unique common perpendicular, which minimizes the distance between the two geodesics; this common perpendicular is thus the curve whose length is $E_W$, labeled as $M$ in Fig.~\ref{Tennis_ball}. The proposal is then that the entanglement of purification can be calculated holographically as proportional to the entanglement wedge cross-section \cite{TU18,Swing18}:
\begin{align}
    E_P(A:B) = \frac{E_W(A:B)}{4G_N}\,.\label{EPEW} 
\end{align}
Interestingly, the primary evidence for this $E_P$-$E_W$ correspondence was monotonicity and the inequalities (\ref{Inequalities}). Given that $E_Q$ and $E_R$ are now known to satisfy the same properties, it is useful to have another way of thinking about the issue, to suggest whether $E_P$ is indeed the correct dual for $E_W$, and to identify the proper gravity dual for the other correlation measures. This involves understanding how an information measure involving an optimization over additional degrees of freedom should be calculated holographically.

Comparing the definition of $E_P$ to the geometry in Fig.~\ref{Tennis_ball} provides a hint. In calculating $E_P$ we are instructed to purify the system $\rho_{AB}$, and then find the partition of the new degrees of freedom into $a$ and $b$ such that $S(Aa)$ is minimized. Consider for concreteness the state $\rho_{AB}$ obtained from the vacuum of the CFT by tracing out other degrees of freedom. Clearly one purification of this is the vacuum state itself, which geometrically corresponds to connecting $A$ and $B$ using the remainder of the boundary circle, denoted $E$ in Fig.~\ref{Tennis_ball}; however there is no reason to believe this is the optimum purification. Looking at the geometry, if we allow ourselves to think of the dashed RT surfaces as spatial regions corresponding to purifications of $\rho_{AB}$, then we can associate $a$ with the left half and $b$ with the right half as indicated in the figure, and then $E_W$ does indeed extremize the area and gives exactly the entropy $S(Aa)$, leading back to the $E_P$-$E_W$ conjecture (\ref{EPEW}). However, for this to make sense we must have a way of associating states not just with boundary regions, but with surfaces in the bulk of the gravity theory.

The surface-state correspondence (SSC) \cite{SurfaceState} is a generalization of AdS/CFT in which a Hilbert space $\mathcal{H}_{tot}$ is associated not only to the AdS boundary, but to surfaces in the interior of the geometry as well.  A pure state on $\mathcal{H}_{tot}$ is dual to a space-like closed convex surface of codimension 2 (an embedding of $S^{n-2}$ into a Cauchy slice $\Sigma$ of AdS$_n$ whose image is convex in $\Sigma$), while a mixed state is dual to a strict subset of a closed convex surface, i.e.~a pure state with a subsystem traced out (erased).  The entropy of a mixed state is given by the area of the smallest surface which closes the dual surface; note that for states whose dual surfaces lie in the AdS boundary, this reduces exactly to the RT formula.  An important implication of this is that a state which is dual to a geodesic in AdS must have no correlation between any of its subsystems; dividing the geodesic into disjoint pieces $a$ and $b$ and applying the entropy rule, one sees that $I(a:b) = 0$.  A heuristic way to view the SSC is that given any space-like closed convex surface $\Omega$ of codimension 2, we have a nested instance of ordinary AdS/CFT, where the ``boundary" is now $\Omega$ and the bulk is its interior.

Using SSC, we can understand better the previous arguments for the conjectured duality of $E_W$ and $E_P$.  Given disjoint regions $A$ and $B$ of the boundary of a Cauchy slice $\Sigma$ of AdS, any curves connecting the boundaries of $A$ and $B$ into a closed convex surface correspond to a purification. To minimize $S(Aa)$ we must choose the purification and partition which minimizes correlations between $a$ and $b$. Since geodesics have no correlation between their subsystems, the correct geometric purification corresponds to the boundary of the entanglement wedge, for which $I(a:b) = 0$ for any partition of this purification. SSC then tells us that $S(Aa)$ is given by the length of the curve $M$, whose minimization over choice of $\{a,b\}$ is $E_W$ by definition.

Tensor network models of AdS/CFT can offer further insight into this process (see for example \cite{Swing12}, \cite{Happy15}). The process of deforming a boundary region $A$ into its bulk RT surface $\Gamma_A$  can be thought of as a process of entanglement distillation, where the degrees of freedom on the boundary are sorted into those entangled with other degrees of freedom (tensors adjacent to the RT surface) and those that are not entangled, and are left behind. The boundary of the entanglement wedge $\partial EW(AB)$ is also the RT surface $\Gamma_E$ for the complement of $A$ and $B$ on the boundary, called $E$ in Fig.~\ref{Tennis_ball}. Thus we can view selecting  $\partial EW(AB)$ as the optimum purification as a process of distilling the minimum set of degrees of freedom needed to provide the entanglement with the degrees of freedom in the entanglement wedge (and hence still represent a purification). 

Several proposals besides $E_P$ have been conjectured for the boundary dual of the entanglement wedge cross section \cite{dutta2019canonical,kudler2019entanglement,tamaoka2019entanglement}.  In \cite{dutta2019canonical}, it was argued that $E_W$ should be dual to half of the \emph{reflected entropy}, an upper bound for $E_P$ that corresponds to a fixed configuration of the optimization in Eq~(\ref{Eq:EPDef}).  In \cite{kudler2019entanglement} the log negativity \cite{vidal2002computable} was conjectured, while in \cite{tamaoka2019entanglement} a quantity called the odd entanglement entropy was proposed to correspond to $E_W$. We do not explore these conjectures further in this paper, but restrict our attention to studying the implications of SSC (and the associated tensor-network models) for more general entropic formulas than the entanglement of purification.  We hope that our findings here can be generalized to give further geometric intuitions from perspectives such as those pursued in \cite{dutta2019canonical,kudler2019entanglement,tamaoka2019entanglement}, but leave such exploration for future work.

In what follows, we will assume the validity of SSC, and apply these methods to calculating holographic duals for $E_Q$ and $E_R$ in the vacuum state of $1+1$-dimensional holographic CFT (our arguments have simple analogues in higher dimensions). We shall see that in this case $E_R$ is also given in terms of $E_W$ using the same formula as $E_P$, Eq. (\ref{EPEW}), while $E_Q$ has a different (but related) holographic realization.

\section{$Q$- and $R$-correlation}

A correlation measure is necessarily monotonically decreasing under local operations.
Optimized correlation measures were studied axiomatically in \cite{Levin19}, and two especially interesting ones found there were the $Q$-correlation and $R$-correlation, which can be defined as minimizations over purifications of the state $\rho_{AB}$ to $\ket{\psi}_{AaBb}$:

\begin{align}
    E_Q(A:B) = \inf_{\ket{\psi}_{AaBb}} f^Q(A,B,a)\quad \quad \quad 
    E_R(A:B) = \inf_{\ket{\psi}_{AaBb}} f^R(A,B,a)
\end{align}
where the functions being minimized are
\begin{align}
    f^Q(A,B,a) &= \frac{1}{2}[S(A) + S(B) + S(Aa) - S(Ba)]\cr 
    f^R(A,B,a) &= \frac{1}{2}[S(AB) + 2S(Aa) - S(ABa) - S(a)]
\end{align}
While we will work with these formulas, it is worth noting we may recast the definitions in terms of conditional entropies,
\begin{align}
    f^Q = \frac{1}{2}[2S(A) + S(a|A) - S(a|B)] \,, \quad \quad \quad 
    f^R = \frac{1}{2}[S(AB) + S(A|a) - S(B|aA)]
\end{align}
or in terms of mutual informations,
\begin{align}
    f^Q = \frac{1}{2}[2S(A) + I(B:a) - I(A:a)] \,, \quad \quad \quad 
    f^R = \frac{1}{2}[2S(A) + I(B:a|A) - I(A:a)]
\end{align}
It also follows from this that the difference between the two is simply half the tripartite information,
\begin{align}
    f^Q - f^R = \frac{1}{2} I_3(A:B:a).
\end{align}
While the tripartite information in general can have either sign, in holographic states one can show that $I_3(A:B:C)\leq 0$, called the monogamy of mutual information (MMI) \cite{hayden2013holographic}. This implies that 
\begin{align}
E_Q(A:B)\leq E_R(A:B)\label{EQER}
\end{align}
for any holographic state $\rho_{AB}$. 

We now use SSC to identify holographic duals of $E_Q$ and $E_R$ in the vacuum state of a two-dimensional CFT, calculating entropies using the generalization of the RT formula to surfaces in the bulk.

\subsection{Conformal invariance of $E_Q$ and $E_R$}

We can argue that in the vacuum state, conformal invariance means it is sufficient to consider a particular subclass of $A$ and $B$ regions without loss of generality. In general, the entanglement entropy of a region in quantum field theory is infinite, and the degrees of freedom must be regulated to obtain a finite result. In the gravity dual, the divergences correspond to the infinite area of surfaces stretching all the way out to the boundary, and so these must be cutoff geometrically. In general $S(A)$ then depends on the cutoff, and conformal invariance is not directly realized.

For certain quantities however, this cutoff dependence cancels. An elementary example is the mutual information of two regions, $I(A:B) = S(A) + S(B)-  S(AB)$. Geometrically, the diverging area of $S(A)$ and $S(B)$ is canceled by that of $S(AB)$, since for each surface headed to the boundary with a plus sign in the RT expression for $I(A:B)$, there is a corresponding surface headed to the same boundary point with a minus sign. As a result, $I(A:B)$ is cutoff independent and hence conformally invariant. Since it is associated to the four points bounding the $A$ and $B$ regions, $I(A:B)$ depends only on the single conformally invariant cross ratio of these four points.

For the quantities we are interested in, the story is the same. Calculations either involve surfaces that do not go to the boundary, and hence are cutoff independent, or go to the boundary at a given point in pairs with opposite signs, so that the cutoff dependence cancels. As a result, once the partition $M$ is chosen, and is allowed to transform under a conformal map via the associated bulk isometry, $f^Q$ and $f^R$ are also conformally invariant and depend only on the cross-ratio of the four defining points.  Therefore $E_Q$ and $E_R$ are also conformally invariant.

For ordered points on the real line, we define the cross-ratio as $(x_2-x_1)(x_4-x_3)/(x_3-x_1)(x_4-x_2)$. Mapping the real line (thought of as the boundary of the upper half plane) to the boundary of the unit disk using the Cayley transform $f(z) = (z-i)/(z+i)$,  the cross ratio in terms of angles on the circle is
\begin{align}
    {\sin (\theta_{12}/2)\, \sin (\theta_{34}/2) \over \sin (\theta_{13}/2)\, \sin (\theta_{24}/2)} \,,
\end{align}
where $\theta_{ij} \equiv \theta_i - \theta_j$.
This cross-ratio can take any value between 0 and 1.

We can obtain a representative configuration for every possible cross-ratio by considering only the ``symmetric" cases where $A$ and $B$ are of equal size with diametrically opposite centers (as in Fig~\ref{Tennis_ball}), in which case the cross-ratio reduces to just $\sin^2(\theta/2)$, with $\theta$ the angular size of each region. Any other configuration of $A$ and $B$ shares its cross-ratio value with one of these cases, and hence $E_Q$ and $E_R$ can be calculated by mapping onto a symmetric case, using a bulk isometry that induces a boundary conformal map. (This would no longer be true for a configuration with less symmetry, like one containing a black hole.) Hence, we will consider the symmetric cases only in what follows. Since the results are cutoff independent, in practice we also assume a spatially uniform cutoff at constant radial coordinate.

\subsection{Geometric duals of $E_Q$ and $E_R$}

Following the arguments in the previous subsection, in what follows we consider only symmetric configurations. In addition we are interested in geometries with connected entanglement wedges, corresponding to $\theta > \pi/2$ for each region. Following the arguments given in the introduction, we assume the optimum purification of $\rho_{AB}$ into $\ket{\psi}_{AaBb}$ corresponds to completing the $A$ and $B$ regions into a closed convex surface by connecting them along the geodesic curve forming the boundary of the entanglement wedge $\partial EW(AB)$. We then leave to be determined how to partition $\partial EW(AB)$ into surfaces $a_1$, $a_2$ associated to $A$ and surfaces $b_1$, $b_2$ associated to $B$, corresponding to a partition of the purifying system into subsystems $\{a,b\}$.  The following proofs are somewhat technical, and the reader interested in the results may find discussion in the next subsection.

Consider first $E_R$. For each term in $f^R$ we must use SSC to find the sum of bulk surfaces dual to the term.  Using the labels in Fig. \ref{Tennis_ball}, $S(AB)$ is always given by $|a_1| + |b_1| + |a_2| + |b_2|$, and $S(Aa)$ is always given by $|M|$.  For $S(ABa)$, there are two options according to SSC: $|b_1| + |b_2|$ or $|\Gamma_B| + |M|$.  Similarly, there are two options for $S(a)$: $|a_1| + |a_2|$ or $|\Gamma_A| + |M|$.  The following fact fixes these options for every partition. 
\begin{prop}\label{prop1}
Given a connected $EW(AB)$ and any partition $\{a,b\}$ of the RT surfaces bounding $EW(AB)$ (as in Fig. \ref{Tennis_ball}), we have $|a_1| + |a_2| < |\Gamma_A| + |M|$ and $|b_1| + |b_2| < |\Gamma_B| + |M|$.
\end{prop}
\begin{proof}
We argue that $|a_1| + |a_2| < |\Gamma_A| + |M|$, and the proof for the other statement is identical.  We will show that
\begin{align}
   |\Gamma_A| + |M| - |a_1| - |a_2| > I(A:B),\label{Sa=1+5}
\end{align}
which will prove the claim since $I(A:B)\geq 0$. Writing $I(A:B)$ as a linear combination of lengths using the RT formula, (\ref{Sa=1+5}) becomes
\begin{align}
    |\Gamma_A| + |M| - |a_1| - |a_2| > |\Gamma_A| + |\Gamma_B| - |a_1| - |b_1| - |a_2| - |b_2|
\end{align}
which is equivalent to 
\begin{align}
    |M| > |\Gamma_B| - |b_1| - |b_2|,
\end{align}
or 
\begin{align}
    |M| + |b_1| + |b_2| > |\Gamma_B| ,
\end{align}
which is true since $\Gamma_B$ is defined to be the minimal area surface connecting the two boundary points of $B$.
\end{proof}
\begin{corr}
Let $\{a,b\}$ be a partition of the RT surfaces bounding $EW(AB)$ (as in Fig. \ref{Tennis_ball}).  Then $I(a:b) = 0$.
\end{corr}
Proposition 1 tells us that in all cases, $S(ABa)$ is given by $|b_1| + |b_2|$ and $S(a)$ is given by $|a_1| + |a_2|$.  Combining all four terms of $f^R$, we are left with only surface $M$, and therefore we conclude that for disjoint and connected boundary regions $A$ and $B$,
\begin{align}
    E_R(A:B) = \frac{E_W(A:B)}{4G_N}.\label{EREW}
\end{align}
Thus for holographic vacuum states, $E_R$ and $E_P$ have the same realization as the entanglement wedge cross-section.  The additivity of $E_R$ \cite{Levin19}, as compared to the conjectured nonadditivity of $E_P$ \cite{NonAdd}, and it's equivalence to $E_W$ in a holographic scenario strengthens the hope of \cite{TU18} that for holographic states $E_P$ is in fact additive.  It may also suggest that $E_R$ should be considered the more fundamental dual of entanglement wedge cross section.  It would be interesting to find holographic states for which these two measures differ.

\begin{figure}
    \centering
    \includegraphics[scale = 0.42, trim=0 156 0 180, clip]{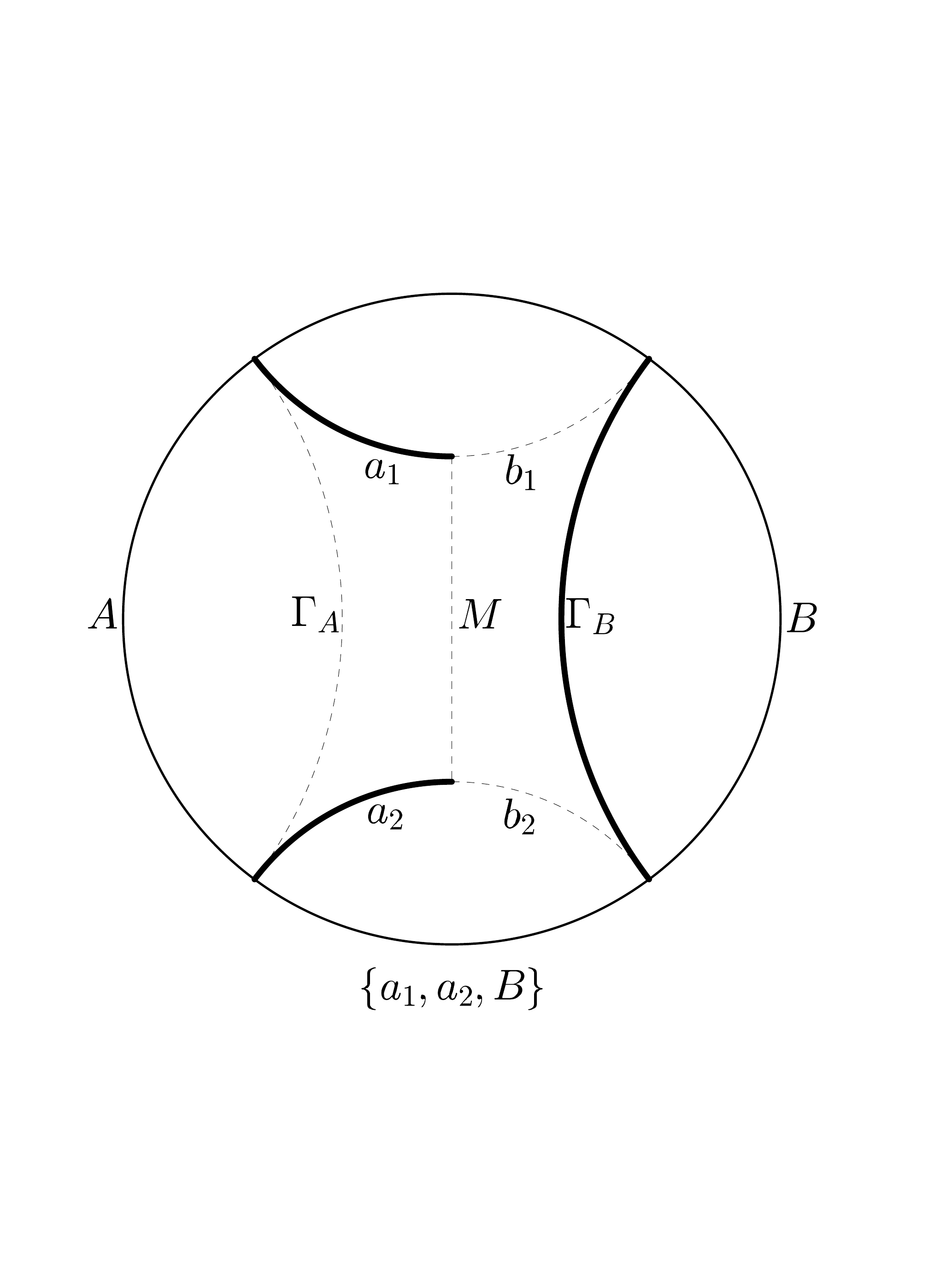}
    \includegraphics[scale = 0.42, trim=0 156 0 180, clip]{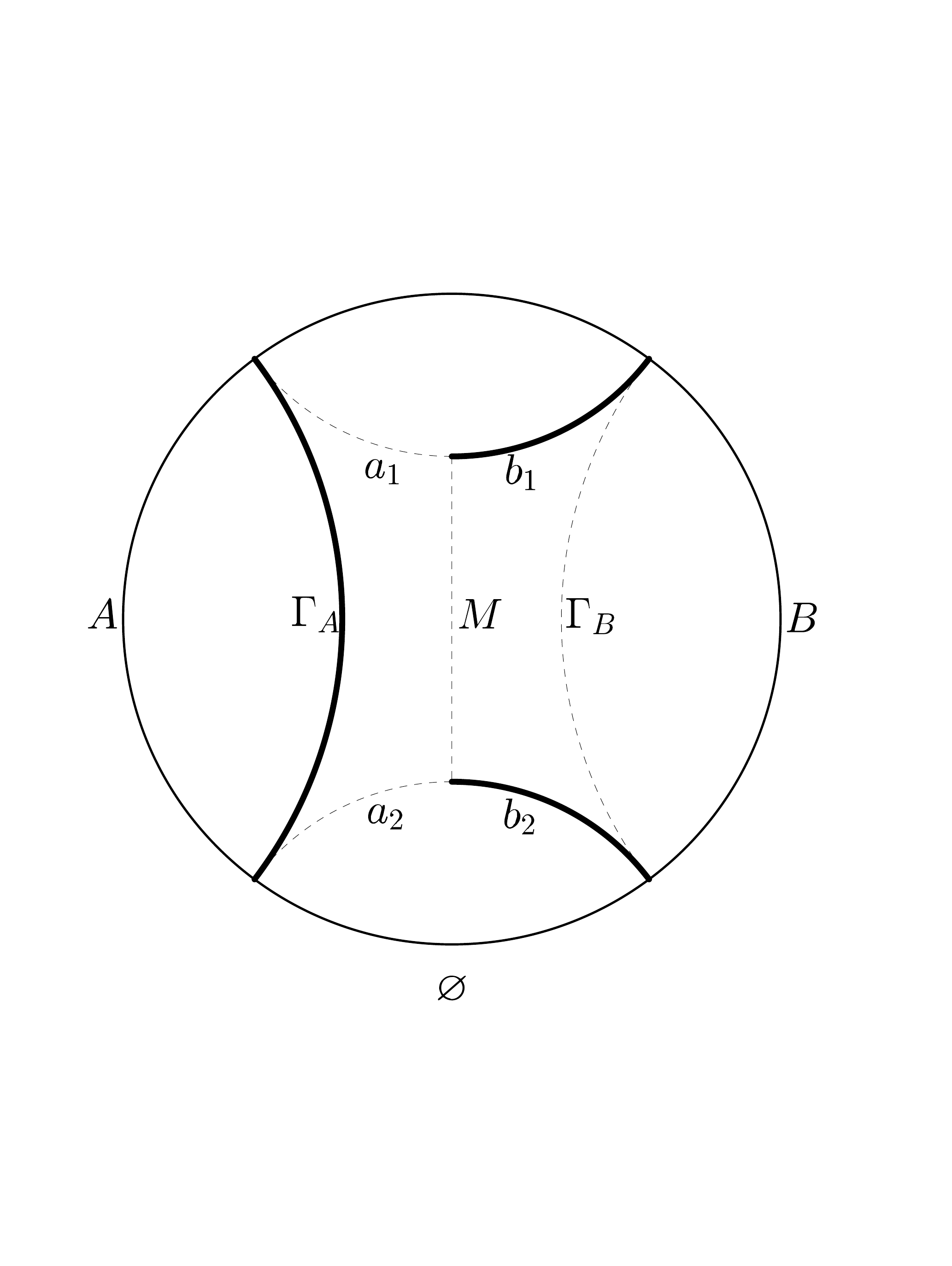}
    \includegraphics[scale = 0.42, trim=0 156 0 150, clip]{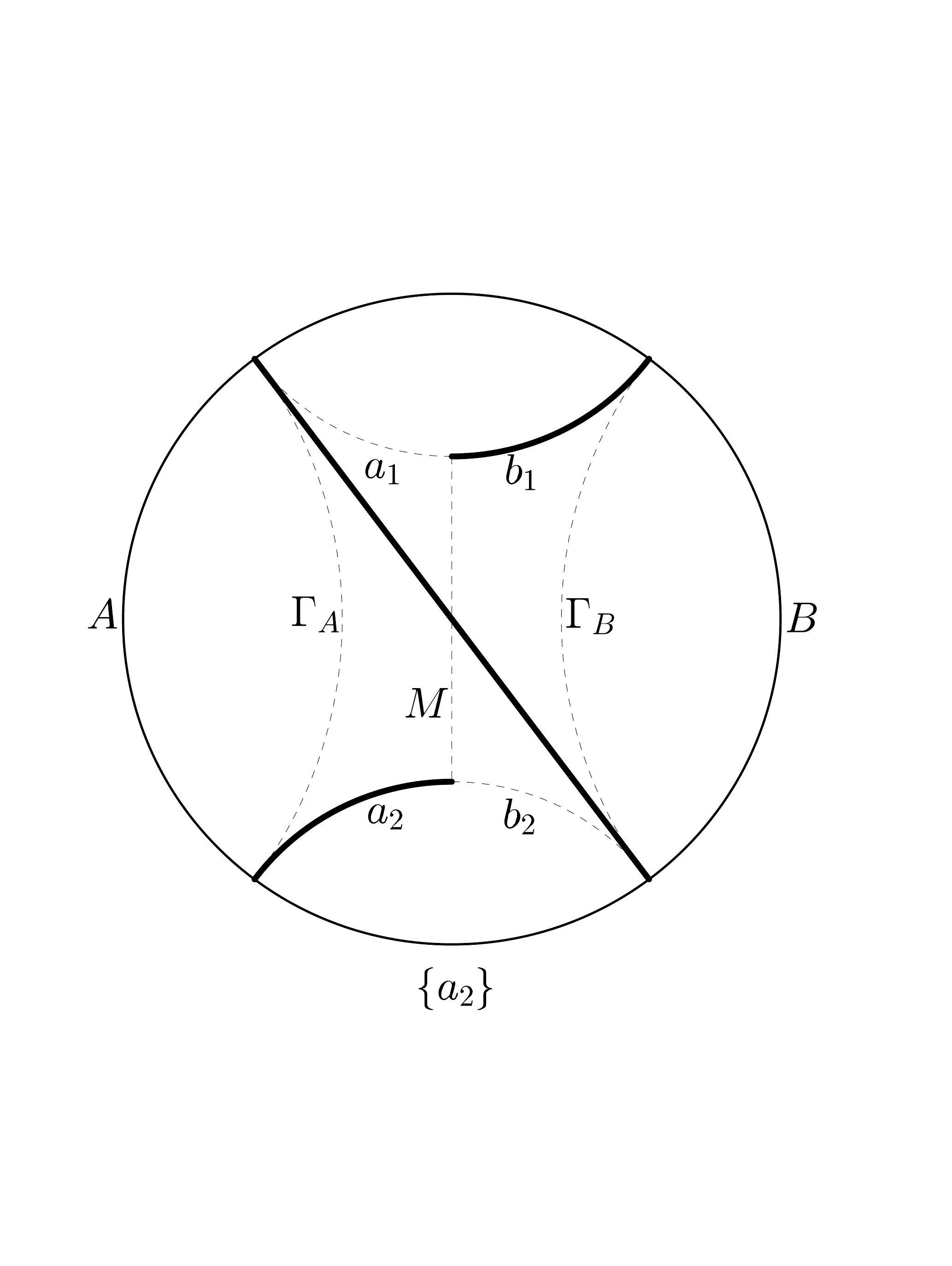}
    \includegraphics[scale = 0.42, trim=0 156 0 150, clip]{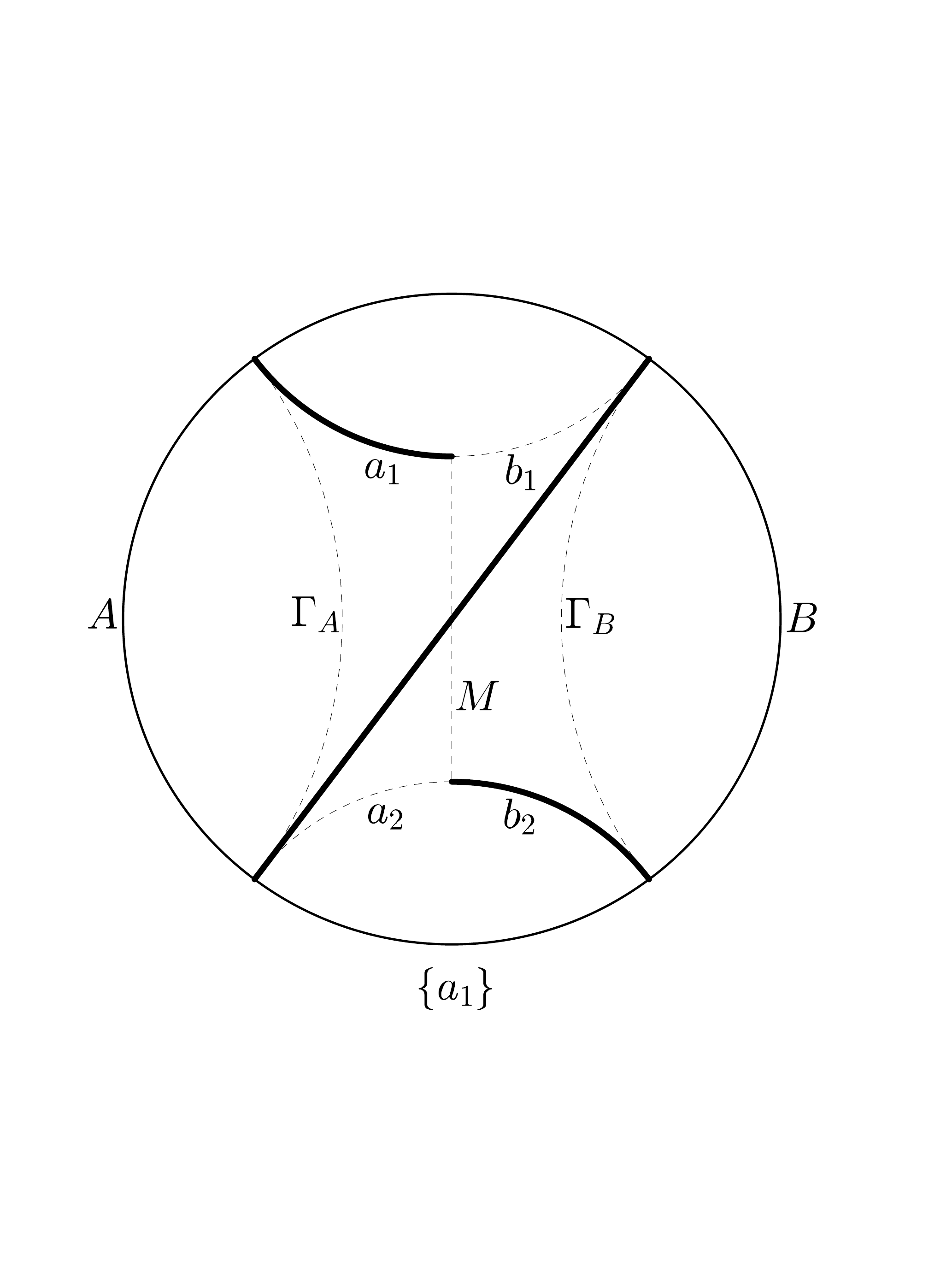}
    \caption{Bulk surface configurations for $S(Ba)$ in each of the four possible cases.}
    \label{S_Ba}
\end{figure}

\begin{figure}
    \centering
    \includegraphics[scale = 0.42, trim=0 156 0 180, clip]{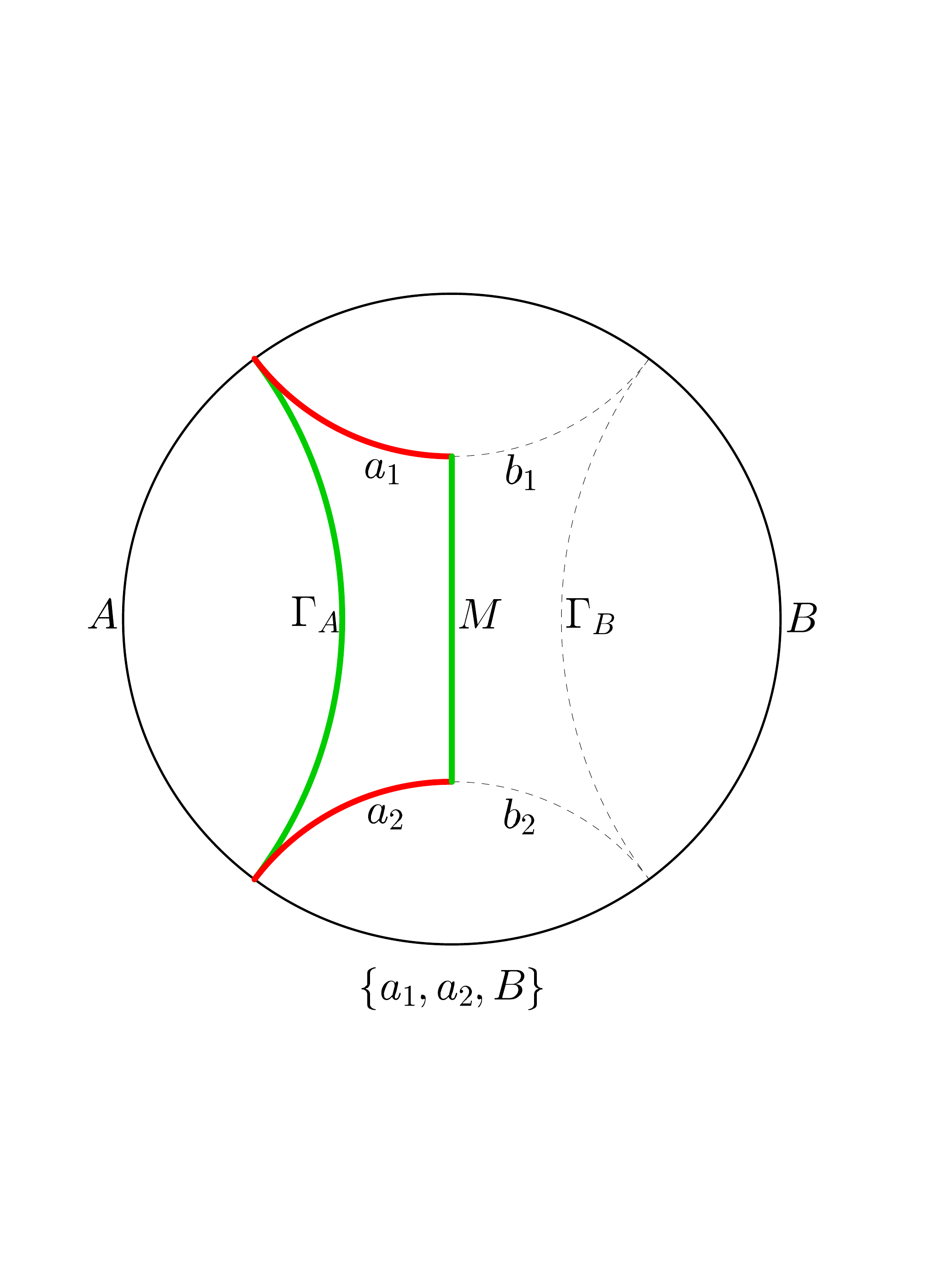}
    \includegraphics[scale = 0.42, trim=0 156 0 180, clip]{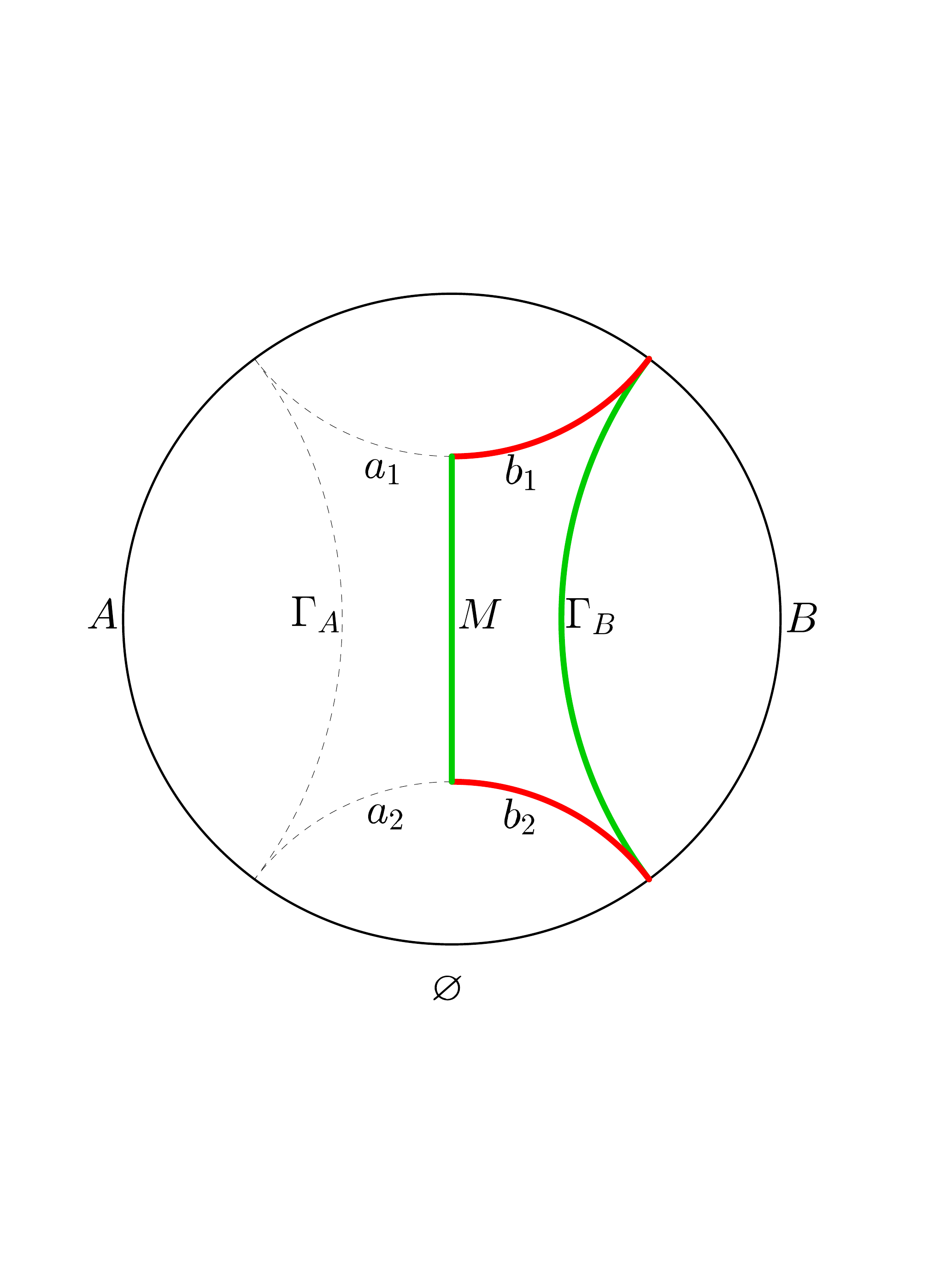}
    \includegraphics[scale = 0.42, trim=0 156 0 150, clip]{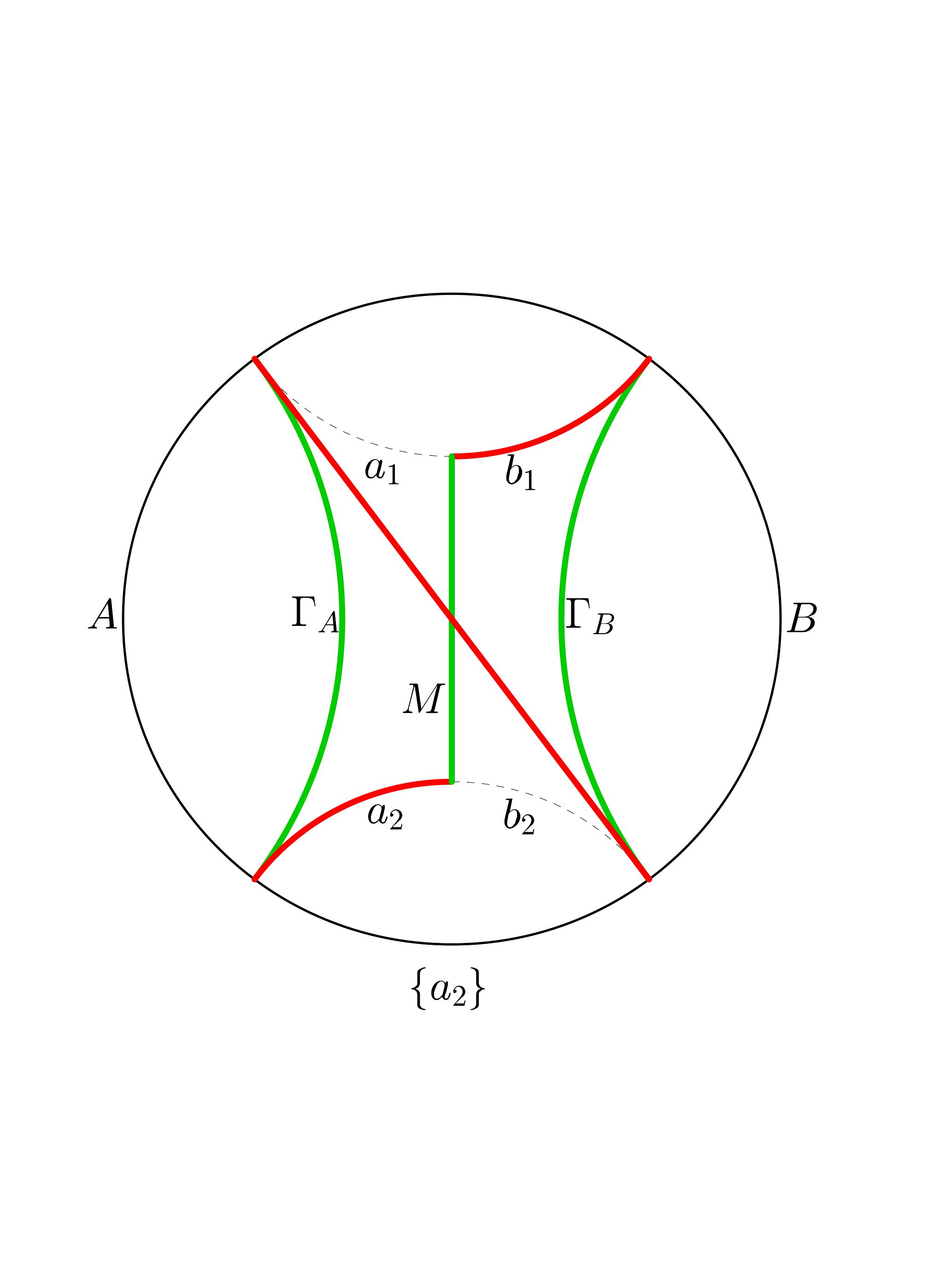}
    \includegraphics[scale = 0.42, trim=0 156 0 150, clip]{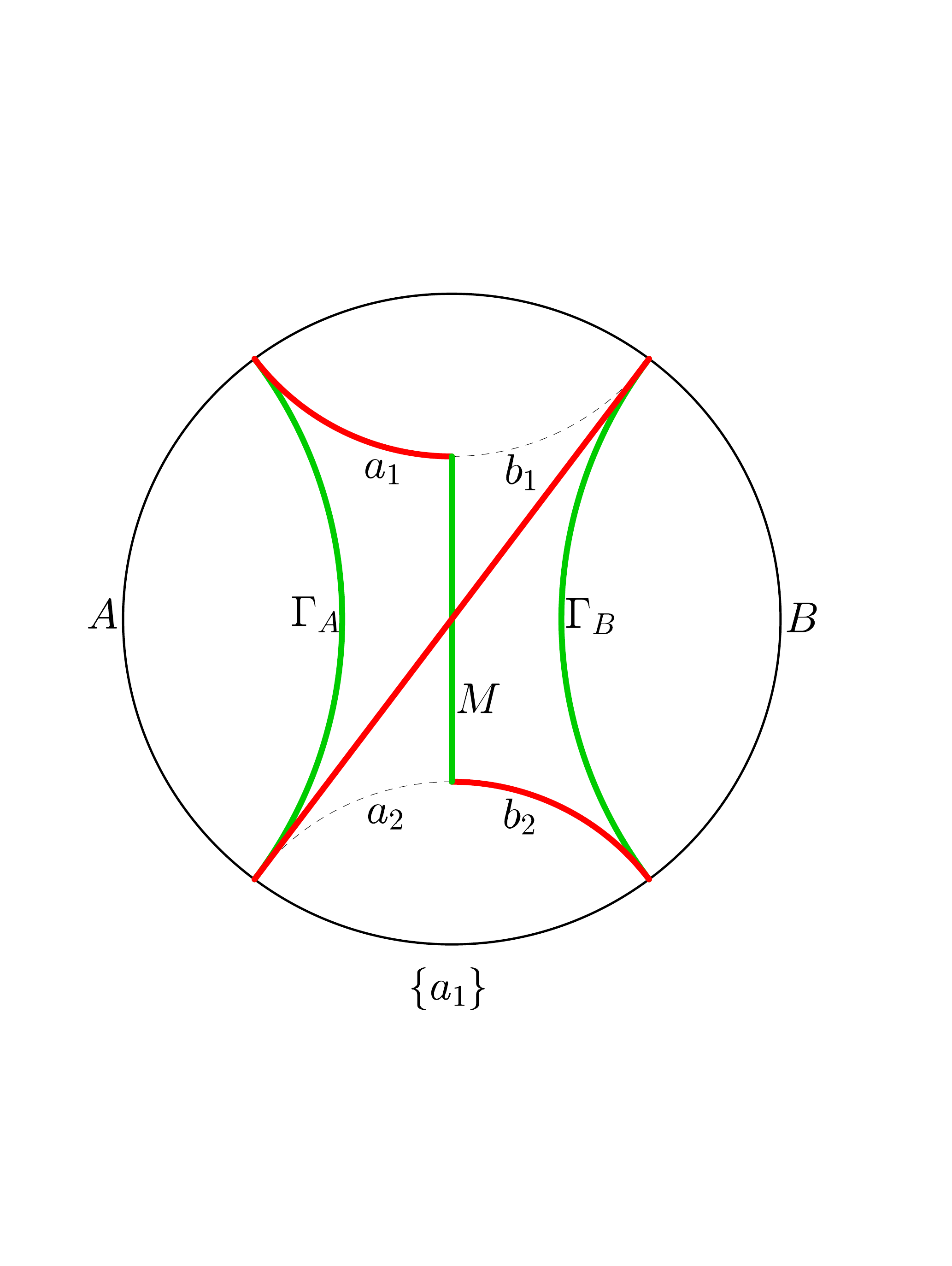}
    \caption{Bulk surface configurations for $f^Q$ in each of the four possible cases.  Green curves are added, red curves are subtracted.  In the order shown, we refer to these quantities as $\Delta A$, $\Delta B$, $\Delta a_2$, $\Delta a_1$.}
    \label{fQ}
\end{figure}

Now we address $E_Q$. We will denote the points partitioning the two components of $\partial EW(AB)$ into $\{a,b\}$ as $p$ and $q$, respectively. For any partition, $S(A)$ and $S(B)$ are always given by their RT surfaces ($|\Gamma_A|$ and $|\Gamma_B|$), and $S(Aa)$ is given by $|M|$.  There are five possible bulk surface configurations for $S(Ba)$.  For the three systems $\{a_1,a_2,B\}$, either all mutual informations are zero, each has positive mutual information with the other two, or one pair has positive mutual information while the pair has zero mutual information with the third. The latter case consists of three subcases: $I(a_1:a_2)>0$, $I(a_1:B)>0$, and $I(a_2:B)>0$.  We will label these five cases by listing all systems in $\{a_1,a_2,B\}$ which have zero mutual information with the other two: $\{a_1,a_2,B\}$, $\varnothing$, $\{B\}$, $\{a_2\}$, and $\{a_1\}$, in the order mentioned in this paragraph.  Prop. \ref{prop1} implies that $\{B\}$ is never the case.  In the four remaining cases, $S(Ba)$ is given by the bulk surfaces shown in Fig. \ref{S_Ba}.  
The linear combinations of bulk surfaces for $f^Q$ in each of the four cases are shown in Fig. \ref{fQ}.
In cases $\{a_1,a_2,B\}$, $\varnothing$, $\{a_2\}$, and $\{a_1\}$, denote by $\Delta A$, $\Delta B$, $\Delta a_2$, and $\Delta a_1$ (respectively) the linear combinations of bulk surfaces which give $f^Q$.  Then, since the only term in $f^Q$ which has multiple options for its bulk surfaces is subtracted (and the correct choice is the minimum option), we have
\begin{align}
    f^Q = \frac{1}{2}\max\{\Delta A, \Delta B, \Delta a_2, \Delta a_1\}
\end{align}
for any partition.  Therefore
\begin{align}
    E_Q(A:B) = \frac{1}{2}\inf_{\ket\psi_{AaBb}}\max\{\Delta A, \Delta B, \Delta a_2, \Delta a_1\}.
\end{align}
Now, if we can show that for any $\{a_2\}$ or $\{a_1\}$ partition, there is always a $\varnothing$ or $\{a_1,a_2,B\}$ partition which achieves a smaller value of $\max\{\Delta A, \Delta B\}$, we will have proven the following:

\begin{prop}\label{prop2}
For disjoint and connected boundary regions $A$ and $B$, 
\begin{align}
E_Q(A:B) = \frac{1}{2}\inf_{\ket\psi_{AaBb}}[\max\{\Delta A, \Delta B\}].\label{E_Q}
\end{align}
\end{prop}

\begin{proof}

We need to show that for any $\{a_1\}$ or $\{a_2\}$ partition $\{p,q\}$ we can find a $\{a_1,a_2,B\}$ or $\varnothing$ partition $\{p^*,q^*\}$ for which 
\begin{align}
    f^Q(\{p^*,q^*\}) \equiv \frac{1}{2}\max\{\Delta A(\{p^*,q^*\}), \Delta B(\{p^*,q^*\})\} \leq \frac{1}{2}\max\{\Delta A(\{p,q\}), \Delta B(\{p,q\})\}
\end{align}
It suffices to prove this for symmetric $A$ and $B$ with a uniform cutoff, since all relevant quantities are cutoff independent\footnote{$f^Q$ satisfies the balancing condition in all cases, and is therefore cutoff independent.  The difference of any two quantities depicted in Fig. \ref{S_Ba} satisfies the balancing condition and is off independent, and therefore the type of partition ($\{a_1,a_2,B\}$, $\varnothing$, $\{a_2\}$, or $\{a_1\}$) is also cutoff independent.}. 

Without loss of generality, let $\{p,q\}$, be a $\{a_2\}$ partition for symmetric boundary regions $A$ and $B$.  Let $p_{EW}$ and $q_{EW}$ be the points on the upper and lower surfaces where $EWCS(AB)$ is anchored (see Fig. \ref{Prop2}).  
\begin{figure}
\begin{minipage}[t]{0.45\linewidth}
\includegraphics[scale = 0.31, trim = 0 0 0 32, clip]{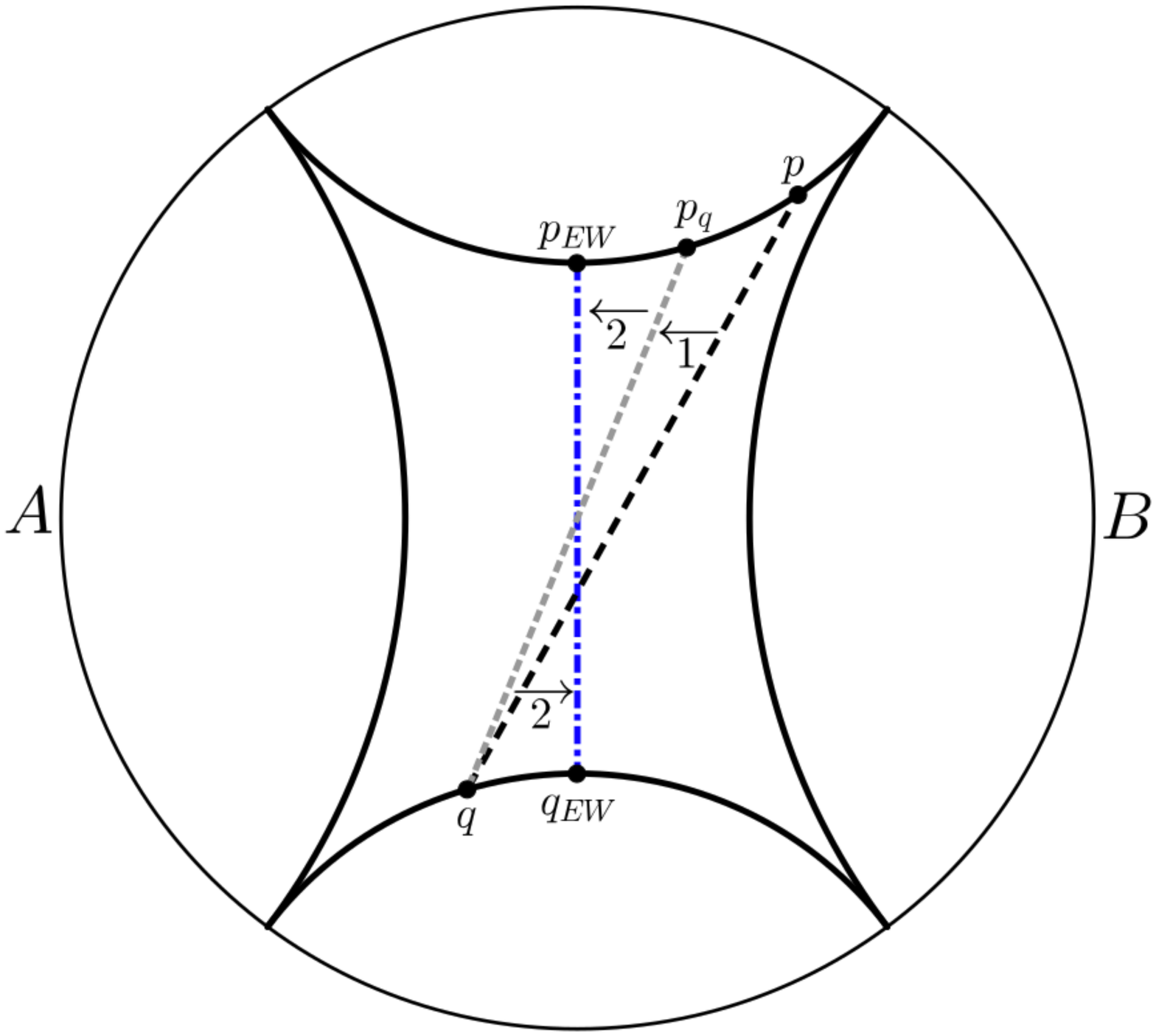}
\caption{Proof of Prop. 2. Starting with an $\{a_2\}$ partition, moves 1 and 2 both decrease the value of $\max\{\Delta A, \Delta B\}$, and the resulting partition is guaranteed to be $\varnothing$ or $\{a_1,a_2,B\}$, which proves the proposition.}\label{Prop2}
\end{minipage}
\hfill
\begin{minipage}[t]{0.45\linewidth}
\includegraphics[scale = 0.39, trim = 0 175 0 160, clip]{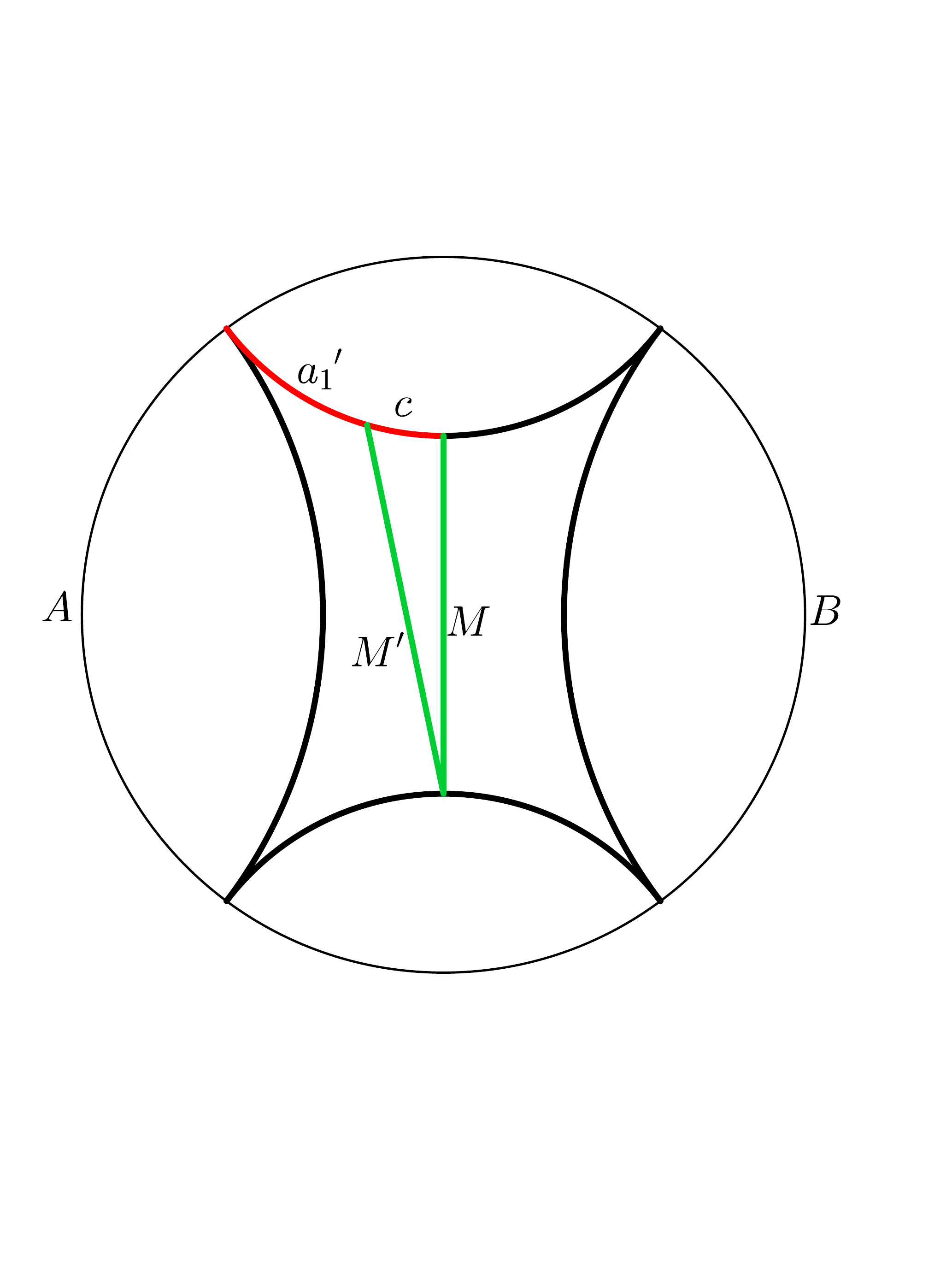}
\caption{When we move the point $p$ towards $A$, the only contributions to $\Delta A$ that change are $|M|$ and $|a_1|$, as shown.  Therefore this move increases $\Delta A$ if and only if $|M|-|c| - |a_1|'<|M'| - |a_1'|$, which is equivalent to $|M|<|M'|+|c|$, which is true by definition since $M$ is a minimal surface.  An identical argument shows that $\Delta B$ decreases.}\label{mv_pt}
\end{minipage}%
\end{figure}
Invoking the symmetry of $A$ and $B$ and the cutoff, it must be the case that $p$ is on the $B$-side of $p_{EW}$ (right, in the figures) and $q$ is on the $A$-side (left).  If we had $p$ on the left and $q$ on the right, then it is clear that $\Delta a_1 > \Delta a_2$, so $\{p,q\}$ cannot be a $\{a_2\}$ partition.  If both $p$ and $q$ are on the right, then looking at Fig. \ref{S_Ba}, it is clear that the $\varnothing$ configuration gives a smaller value for $S(Ba)$ than the $\{a_2\}$ configuration (again by symmetry, and also the fact that a diameter is the longest possible geodesic given a uniform cutoff), so again $\{p,q\}$ cannot be a $\{a_2\}$ partition.  Similarly, with both on the left, the $\{a_1,a_2,B\}$ configuration gives a smaller value of $S(Ba)$ than the $\{a_2\}$ configuration.  

Without loss of generality, assume that $d(p,p_{EW}) \geq d(q,q_{EW})$.  Then it is clear from symmetry that $\Delta B \geq \Delta A$, since $|b_2|\leq |a_1|$ and $|b_1|\leq |a_2|$.  Now, consider what happens to $\Delta A$ and $\Delta B$ as we move $p$ towards $A$.  Regardless of the initial partition, this will always increase $\Delta A$ and decrease $\Delta B$, as Fig. \ref{mv_pt} shows.

If we continue moving point $p$ towards $A$ we will reach a point $p_q$ where $\Delta A = \Delta B$ (this is move 1 in Fig. \ref{Prop2}).  By symmetry, $d(p_q,p_{EW}) = d(q,q_{EW})$.  Since move 1 has decreased $\Delta B$, $\max\{\Delta A, \Delta B\}$ has also decreased.  If the resulting partition is a $\varnothing$ or $\{a_1,a_2,B\}$ partition\footnote{Actually, since $\Delta A = \Delta B$, if $\Delta A > \Delta a_1,\Delta a_2$ then the partition lies on the boundary between $\varnothing$ and $\{a_1,a_2,B\}$ partitions.} then we are done.  If it is not, then move $p$ to $p_{EW}$ and move $q$ to $q_{EW}$ (this is move 2 in Fig. \ref{Prop2}).  The $\{p_{EW},q_{EW}\}$ partition is certainly a $\varnothing$ or $\{a_1,a_2,B\}$ partition, since looking at Fig. \ref{S_Ba} again we can see that the $\varnothing$ and $\{a_1,a_2,B\}$ configurations both give smaller values for $S(Ba)$ than both the $\{a_1\}$ and $\{a_2\}$ configurations (again, since diameters are the longest geodesics given a uniform cutoff).  Furthermore, the only change to the surfaces of $\Delta A$ and $\Delta B$ which does not cancel out is the change in $|M|$, which has decreased since we moved to $EWCS(AB)$.  Therefore, in move 2, $\Delta A$ and $\Delta B$ have both decreased (by exactly the decrease in $|M|$), and we end in a $\varnothing$ or $\{a_1,a_2,B\}$ partition.  This proves the proposition.
\end{proof}

\subsection{Properties of the $Q$-correlation}

Prop. \ref{prop2} gives us a geometric interpretation of the $Q$-correlation: it is given by the surfaces in the top left (or equivalently top right) image of Fig. \ref{fQ}, by adding the RT surface $\Gamma_A$ giving the entropy for $A$ and the cross-section $M$ bounding $Aa$, and subtracting the surface for $a$ itself. Similarly, Eq. (\ref{EREW}) gives us a geometric interpretation of the $R$-correlation: it is given by $E_W$.  These dual descriptions allow us to geometrically verify the inequalities (\ref{Inequalities}) for $E_Q$ (this has been done already for $E_W (= E_R)$ in \cite{TU18}).  To see that $E_Q(A:B)\geq \frac{1}{2}I(A:B)$, note that $\Delta A = \Delta B$ for the optimal partition implies that 
\begin{align}
|\Gamma_A| - |a_1| - |a_2| = |\Gamma_B| - |b_1| - |b_2| = \frac{1}{2}I(A:B).  
\end{align}
Therefore
\begin{align}
    E_Q(A:B) - \frac{1}{2}I(A:B) = \frac{1}{2}(|M| + |a_1| + |a_2| - |\Gamma_A| ),
\end{align}
and the right side is a positive quantity since $|M| + |a_1| + |a_2|\geq |\Gamma_A|$ is evident from the minimality of $|\Gamma_A|$.  To see that $E_Q(A:B)\leq \min\{S(A),S(B)\}$, observe that 
\begin{align}
    E_Q(A:B) - S(A) = \frac{1}{2}(|M| - |\Gamma_A| - |a_1| - |a_2|),
\end{align}
and the right side is again a positive quantity, due to the minimality of $|M|$.  Finally, the monogamy inequality in (\ref{Inequalities}), for both $E_Q$ and $E_R$, follows from their common lower bound of $\frac{1}{2}I(A:B)$ and MMI. Monotonicity under local processing was shown geometrically for $E_W (= E_R)$ in \cite{TU18}, and for $E_Q$ in \cite{umemoto2019quantum}.

We can also see that the $Q$-correlation is achieved by the same partition as $E_W$:
\begin{corr}
For disjoint and connected boundary regions $A$ and $B$, $E_Q(A:B) = f^Q(\{p_{EW},q_{EW}\})$, i.e. $E_Q$ is achieved at $EWCS(AB)$.
\end{corr}

\begin{proof}
Again, it suffices to prove this for the case of symmetric boundary regions and a uniform cutoff, since all relevant quantities are cutoff independent.  Fig. \ref{mv_pt} tells us that $\Delta A$ and $\Delta B$ both have no local minima.  Therefore the form of (\ref{E_Q}) implies that $E_Q$ is achieved by a partition where $\Delta A = \Delta B$.  In the symmetric case, these partitions are the ones for which $p$ and $q$ are on opposite sides of $EWCS(AB)$, with $d(p,p_{EW}) = d(q,q_{EW})$.  We can move from any one of these partitions to any other by simply moving $p$ and $q$ in opposite directions the same distance.  Like in the proof of Prop. 2, this move only changes $M$'s contribution to $\Delta A$ and $\Delta B$, since all other changes cancel out.  Therefore the partition with $\Delta A = \Delta B$ which achieves the mininum value of $\Delta A$ and $\Delta B$ is the one with minimal $|M|$.  Since the $EWCS(AB)$ partition clearly has the property that $\Delta A = \Delta B$, and has the minimal $|M|$ by definition, we conclude that this partition achieves $E_Q$.
\end{proof} 

Thus the minimizations leading to $E_W$, $E_Q$ and $E_R$ all involve the same purification and the same partition into $\{a, b\}$.  This now allows us to geometrically verify inequality (\ref{EQER}).  To see that $E_Q(A:B)\leq E_R(A:B)$, note that 
\begin{align}
    E_Q(A:B) - E_R(A:B) = \frac{1}{2}(|\Gamma_A| - |a_1| - |a_2| - |M|),
\end{align}
and the right side is a negative quantity, again due to the minimality of $|\Gamma_A|$.

We can think of the optimal partition as ensuring that $a$ is as uncorrelated with $B$ as possible. In particular, we can prove that for the optimal partition, $I(a:B)  = I(b:A) = 0$:

\begin{corr}\label{IaB}
For disjoint and connected boundary regions $A$ and $B$ in the optimal partition, $I(a:B) = I(b:A) = 0$.
\end{corr}
\begin{proof}
Since the partition is not $\{a_1\}$ or $\{a_2\}$, we know that both $S(Ba)$ and $S(Ab)$ are given by either $|\Gamma_A| + |b_1| + |b_2|$ or $|\Gamma_B| + |a_1| + |a_2|$.  But since $\Delta A = \Delta B$, we know that $|\Gamma_A| + |b_1| + |b_2| = |\Gamma_B| + |a_1| + |a_2|$.  Observing that $S(B) + S(a)$ is given by $|\Gamma_B| + |a_1| + |a_2|$ and $S(A) + S(b)$ is given by $|\Gamma_A| + |b_1| + |b_2|$, the  claim follows.
\end{proof}

Furthermore, this leads to another way to understand the value of $E_Q$: it is proportional to the mutual information $I(Aa:B)$ between the $Aa$ system and $B$, after all correlation between both $a$ and $B$, and $b$ and $A$, has been eliminated:
\begin{corr}\label{IaB2}
For disjoint and connected boundary regions $A$ and $B$, in the optimal partition, $I(Aa:B) = I(Bb:A) = 2 E_Q(A:B)$
\end{corr}
\begin{proof}
We simply observe that $I(Aa:B) = \Delta B$ and $I(Bb:A) = \Delta A$ for any partition, since $|M|$ gives both $S(Aa)$ and $S(Bb)$, $S(B)$ and $S(A)$ are given by their RT surfaces $\Gamma_B$ and $\Gamma_A$, and $S(ABa)$ and $S(ABb)$ are given (using Prop. \ref{prop1}) by $|b_1| + |b_2|$ and $|a_1| + |a_2|$.
\end{proof}
Using this, another interesting presentation of $E_Q$ follows from SSC.
Another important feature of SSC is the following condition for unitary equivalence. If two surfaces representing mixed states share the same boundary, with no extremal surface lying between them (e.g. $A$ and $\Gamma_A$ in Fig. \ref{Tennis_ball}) then their dual states differ by a unitary, serving to concentrate the entanglement as we move closer to a geodesic sharing their boundary.  This is realized by the entropy formula in SSC, since all such surfaces are closed by the same extremal curve and therefore have the same entropy. Now, interpreting $M$ as a system which purifies $Aa$ or $Bb$, we have $\Delta A = I(A:M)$ and $\Delta B = I(B:M)$, so
\begin{align}
    E_Q(A:B) = \frac{1}{2}\inf_M~[\max\{I(A:M),I(B:M)\}].
\end{align}
Since $\rho_M$ differs from both $\rho_{Aa}$ and $\rho_{Bb}$ by a unitary, we have $I(A:M) = I(A:Bb)$ and $I(B:M) = I(B:Aa)$, which again gives the claim of Corr. \ref{IaB2}. Thus we my think about $E_Q$ as a minimal mutual information between a surface region $A$ and a cross-section of $A$'s entanglement wedge with $B$.


\section{Relationship Between $Q$-correlation and Symmetric Side-channel Distillable Entanglement}

The symmetric side-channel assisted distillable entanglement of a state \cite{SSW06}, denoted $I^{ss}(A\rangle B)$, is a measure of bipartite quantum correlation defined by
\begin{align}
   I^{ss}(A\rangle B) & = \frac{1}{2}\sup_{U:A\rightarrow ab}[-S(a|Bb) +S(a|Eb)]
\end{align}
where $E$ purifies $AB$.  Here $U:A \to ab$ represents an isometry from $\mathcal{H}_A$ to a Hilbert space $\mathcal{H}_a\otimes\mathcal{H}_b$. $I^{ss}(A\rangle B)$ can be interpreted operationally as the amount of the entanglement Alice and Bob can distill from their shared state if they have a symmetric channel from Alice to Bob. Symmetric side-channels are stronger than classical communications, and thus $I^{ss}(A \rangle B)$ is an upper bound on the one-way distillable entanglement, $D^{\rightarrow}(\rho_{AB})$.  $I^{ss}(A \rangle B)$ is much better behaved than one-way distillable entanglement: it is convex and additive, while one-way distillable entanglement is neither.  It has also proven a useful tool for finding upper bounds on $D^{\rightarrow}(\rho_{AB})$ \cite{SS08}.

\subsection{Entanglement with Eve vs $Q$-Correlation with Bob}
We find the following relationship between the correlation measures $E_Q$ and $I^{ss}$, independent of holography:
\begin{prop}\label{Iss}
Given a state on $A$ and $B$, and an environment $E$ which purifies the state,
\begin{align}
S(A) & = E_Q(A:B) + I^{ss}(E\rangle A).\label{Iss_EQ}
\end{align}
\end{prop}

\begin{proof}
\begin{align}
I^{ss}(E\rangle A) & = \frac{1}{2}\sup_{E\rightarrow ea}[-S(e|Aa)+ S(e|Ba)]\nonumber  \\
& = \frac{1}{2}\sup_{E\rightarrow ea}\left[-S(eAa) + S(Aa)+ S(eBa)-S(Ba)\right]\nonumber\\
& = \frac{1}{2}\sup_{E\rightarrow ea}\left[ - S(B) + S(Aa) + S(A) - S(Ba)\right]\nonumber\\
& = S(A)+ \frac{1}{2}\sup_{E\rightarrow ea}\left[ - S(B) + S(Aa) -S(A) - S(Ba)\right]\nonumber\\
& = S(A)+ -\frac{1}{2}\inf_{E\rightarrow ea}\left[  S(B) - S(Aa) +S(A) + S(Ba)\right]\nonumber\\
& = S(A) - E_Q(A:B)\nonumber.
\end{align}
\end{proof}

Notice that since $S(A)$ is fixed, Eq. (\ref{Iss_EQ}) implies that the partition which achieves $E_Q(A:B)$ also achieves $I^{ss}(E\rangle A)$. This equation tells us that the entropy of Alice's system can be decomposed into two contributions: $Q$-correlation with Bob, plus entanglement with Eve that can be distilled via a symmetric channel from Eve to Alice. A similar relationship relating $E_P$ and the dense coding advantage was noted in \cite{HP12}.  Eq. (\ref{Iss_EQ}) is also reminiscent the decomposition of \cite{koashi2004monogamy},
\begin{align}
    S(A) = E_f(A:B) + I^{\leftarrow}(A:E),
\end{align}
which embodies the monogamy of entanglement. 
In this equation $E_f(A:B)$ is the entanglement of formation of $\rho_{AB}$ and $I^{\leftarrow}(A:E)$ is the Holevo information of $A$ with respect to $E$, maximized over POVMs on $E$.  Eq.~(\ref{Iss_EQ}) further clarifies the relationship between $A$'s distillable entanglement with $E$ and $A$'s correlation with $B$.  

\subsection{Entanglement, $Q$-Correlation, and the Entanglement Wedge}

As a corollary of Prop. 3, we have the following expression for $I^{ss}(A\rangle B)$ for a pure holographic state on $ABE$:
\begin{corr}
If $\rho_{ABE}$ is a pure holographic state with $A$ and $B$ disjoint and connected, and $\rho_M$ is a state dual to $EWCS(AB)$ via SSC (with $ab$ the purification of $AB$ determined by $M$, as in Fig. \ref{Tennis_ball}), then
\begin{align}
    I^{ss}(E\rangle A) = \frac{1}{2}I(A:a).\label{Iss_IAa}
\end{align}
\end{corr}
\begin{proof}
\begin{align}
    I^{ss}(E\rangle A) &= S(A) - \frac{1}{2}I(A:M)\nonumber
    \\&= \frac{1}{2}[S(A) + S(A|M)]\nonumber
    \\&= \frac{1}{2}[S(A) - S(A|a)]\nonumber
    \\&= \frac{1}{2}I(A:a)\nonumber
\end{align}
\end{proof}
This can also be seen by simply constructing $I(A:a)$ out of bulk surfaces, and noting that $2S(A) - \Delta A = I(A:a)$.  If $I^{ss}(E\rangle A) = \frac{1}{2}I(A:a)$ for some $U:E\rightarrow ab$, then $\rho_{aB} = \rho_a \otimes \rho_B$.  To see this, note that for an optimal\footnote{By optimal we mean a partition which achieves both $I^{ss}(E\rangle A)$ and $E_Q(A:B)$, which exists by Prop. \ref{Iss}.} state $\ket{\psi}_{AaBb}$, we have
\begin{align}
    I^{ss}(E\rangle A) & = \frac{1}{2}[-S(a|Ab)+ S(a|Bb)]\nonumber\\ 
 & = \frac{1}{2}I(A:a)\nonumber,
\end{align}
which is equivalent (because $S(a|Bb)= -S(a|A)$) to 
\begin{align}
-S(a|Ab) -S(a|A) & = S(a) - S(a|A).\nonumber
\end{align}
In turn, this is equivalent to 
\begin{align}
    0 & = S(a) + S(a|Ab)\nonumber\\
    & = S(a) - S(a|B) = I(a:B).\nonumber
\end{align}
Zero mutual information is exactly the condition for $\rho_{aB} = \rho_a\otimes \rho_B$.  This is nicely consistent with Corr. \ref{IaB}.  It is also interesting to note that Eq. (\ref{Iss_IAa}) gives us a geometric dual for $I^{ss}$. Since $E_Q(A:B)$ and $I^{ss}(E\rangle A)$ are achieved by the same purification $\ket{\psi}_{AaBb}$, we can simply read off the RT-surfaces for $I^{ss}(E\rangle A)$ from the right side of Eq. (\ref{Iss_IAa}).  We find that $I^{ss}(E\rangle A)$ is given in bulk surfaces by $\frac{1}{2}(|\Gamma_A|+|a_1|+|a_2|-|M|)$.  Note that this is a cutoff dependent quantity since it is not balanced at $\partial A$.  The cutoff dependence of $I^{ss}(E\rangle A)$ is also evident from Eq. (\ref{Iss_EQ}), since the left side is cutoff dependent and $E_Q$ is cutoff independent.

\section{Discussion}

In this paper we have used SSC to identify the holographic duals of the optimized correlation measures $E_Q$ and $E_R$ introduced in \cite{Levin19}.  We then used these geometric duals to show several properties of $E_Q$ and $E_R$, including inequalities which are known to hold for these correlation measures independent of geometry.  We have also shown a tradeoff relationship between $E_Q(A:B)$ and the symmetric side-channel assisted distillable entanglement $I^{ss}(E\rangle A)$, where $E$ purifies $AB$, which holds for general quantum states.  This relationship allowed us to construct a holographic dual for $I^{ss}$ as well.  We have also argued that for a holographic state, the same purification and partition achieves $E_W(A:B)$, $E_Q(A:B)$, $E_R(A:B)$, and $I^{ss}(E\rangle A)$. For this we used the cutoff independence of $E_Q$, and the existence of a boundary conformal map (bulk isometry) taking any disjoint and connected boundary regions $A$ and $B$ to a pair of symmetric boundary regions.   

We believe that the tradeoff between $E_Q$ and $I^{ss}$ embodied in Eq.~(\ref{Iss_EQ}) will find a use beyond the current setting.  It tells us that the symmetric side-channel distillable entanglement from $E$ to $A$ consists of exactly the correlations that are not captured as $Q$-correlations with $B$.  This may lead us to a deeper understanding of distillability in general and $I^{ss}$ in particular.  It may also point towards an operational understanding of $E_Q$.  We also note that $I^{ss}$ can be interpreted as the utility of a state when used as a quantum one-time pad \cite{BO12a,BO12b}, pointing towards the interpretation of $E_Q(A:B)$ as information that is ``public'' from the perspective of $E$. 

While the Ryu-Takayanagi formula for holographic entanglement entropy is by now well established, the identification of surfaces in the bulk with states having a particular entropy is much less well-tested. The coherence of our results, including the fact that the inequalities satisfied by $E_Q$ and $E_R$ independent of geometry can also be shown to be satisfied geometrically by their duals obtained via the SSC prescription, provides further evidence for this correspondence. Furthermore, it suggests a broader context for the holographic realization of information measures involving the optimization over a purification. It would be interesting to explore this further.

In this work we have focused on the simplest holographic spacetime of pure AdS space, dual to the field theory vacuum. It would be interesting to consider the holographic realization of these correlation measures in more general backgrounds, such as black hole geometries.  Another interesting avenue to explore is a reformulation of our findings in the language of bit threads, introduced in \cite{freedman2017bit} and discussed in the context of $E_W$ in \citep{Bao_BTEP_19,Du19,agon2019, harper2019bit}.  The bit threads formalism may provide an intuition which connects the ``flow" of information from one boundary region to another to our findings about one-way distillable entanglement in holography.

\emph{Note Added:} In the latter stages of this work, we learned of an independent effort by Koji Umemoto which reaches partially overlapping conclusions \cite{umemoto2019quantum}.

\emph{Acknowledgements:} We are grateful to Daniel Hackett, Ethan Neil, Michael Perlin, Markus Pflaum, and Michael Walter for interesting discussions and helpful advice. We also thank Koji Umemoto for making us aware of his work prior to publication.
JL and GS were partially supported by NSF CAREER award CCF 1652560.
OD is supported by the Department of Energy under grant DE-SC0010005.  This work was supported by the Department of Energy under grant DE-SC0020386.

\bibliography{HolCor}

\end{document}